\documentclass[multphys,vecphys]{svmult}

\usepackage{makeidx}         % allows index generation
\usepackage{graphicx}        % standard LaTeX graphics tool
\usepackage{multicol}        % used for the two-column index
\usepackage[bottom]{footmisc}% places footnotes at page bottom
\makeindex             % used for the subject index
\begin{document}

\title*{Enigma of ultraluminous X-ray sources may be resolved by 
3D-spectroscopy (MPFS data)
}
\titlerunning{Enigma of ULXs may be resolved by 3D-spectroscopy}
\author{S. Fabrika and P. Abolmasov
}
\institute{Special Astrophysical Observatory, Russia\\
\texttt{fabrika@sao.ru}
}
\maketitle

\begin{abstract}
The ultraluminous X-ray sources (ULXs) were isolated in external galaxies 
for the last 5 years. Their X-ray luminosities exceed 100-10000 times 
those of brightest Milky Way black hole binaries and they are extremely 
variable. There are two models for the ULXs, the best black hole candidates. 
1. They are supercritical accretion disks around a stellar mass black hole 
like that in SS433, observed close to the disk axes. 2. They are 
Intermediate Mass Black Holes (of 100-10000 solar masses). Critical 
observations which may throw light upon the ULXs nature come from 
observations of nebulae around the ULXs. We present results of 
3D-spectroscopy of nebulae around several ULXs located in galaxies at 
3-6 Mpc distances. We found that the nebulae to be powered by their 
central black holes. The nebulae are shocked and dynamically perturbed 
probably by jets. The nebulae are compared with SS433 nebula (W50).
\end{abstract}

\section*{Introduction}

The main properties of the ultraluminous X-ray sources (ULXs)~--
huge luminosities ($10^{39-41}\,erg/s$), diversity of X-ray spectra,
strong variability, connection with star-forming regions, their 
surrounding nebulae. ULXs may be supercritical accretion
disks observed close to the disk axis in close binaries with a stellar mass
black hole or microquasars (\cite{FaMe01}, \cite{Kingetal01},
\cite{Koerdetal01}). Another idea is that ULXs may be intermediate-mass 
black holes (IMBHs) with "normal" accretion disks
(\cite{CoMu99}, \cite{Milletal04}). It is also possible that ULXs
are not homogeneous class of objects. 

It was suggested originally by Katz \cite{Katz87} 
that SS433 being observed 
close to the jet axis, will be extremely bright X-ray source. 
Fabrika \& Mescheryakov \cite{FaMe01} 
discussed  observational properties of face-on SS433-like objects and 
concluded that they may appear as a new type of extragalactic X-ray sources. 
In \cite{FaKaAbSh06} we discussed possible properties of the funnel in the 
supercritical accretion disk of SS433. We predicted X-ray spectra 
and temporal behaviour of the funnel in "face-on SS433" star in 
application for ULXs. Here we continue to develop this idea and consider 
nebulae surrounding the ULXs sources.

The main difference between SS433 and other known X-ray binaries is
highly supercritical and persistent mass accretion rate
($\sim 10^{-4}\,M_{\odot}/y$) onto the relativistic star (a probable
black hole, $\sim 10 M_{\odot}$), which has led to the formation of a
supercritical accretion disk and the relativistic jets. SS433 properties
were reviewed recently by \cite{Fab04}.

Similar to SS\,433 the ULXs are connected with nebulae. 
They are frequently located in bubble-like nebulae. New data \cite{Pakull06}
show that the nebulae are expanding with a velocity $\sim 80 \,km/s$
(up to $\sim 250 \,km/s$). The nebula sizes are from 20 to a few hundred 
parsecs, such nebulae are easy for observations even from  
"megaparsec distances". Here we compare the gas nebula around
SS\,433 with nebulae of ULXs in Holmberg II, NGC\,6946 and IC\,342 galaxies 
observed recently by the Integral-field spectroscopy methods.

\section*{Observations and data reduction}
\label{sec:1}

We discuss results of observations on the Russian 6-m telescope
with the integral field spectrograph MPFS
\cite{Afanetal01}.
The integral field unit of 16$\times$16 square spatial elements
covers a region of 16"$\times$16" ~on the sky.
Integral field spectra were taken in
the spectral range 4000~-- 6800~\AA\AA\ with a seeing 1.0-1.3" (FWHM). 
Data reduction was made using procedures developed in IDL environment 
(version 6.0) by V.\,Afanasiev, A.\,Moiseev and P.\,Abolmasov and 
include all the standard steps.

\section*{Nebulae surrounding ULXs and SS433}

The object SS\,433 is surrounded by the elongated radio nebula W\,50, which 
was produced (or distorted) by SS433 jets due to the jet interaction 
with interstellar medium \cite{Dubetal98}. The radio emission is 
synchrotron, relativistic electrons appear at the jet deceleration.
Bright optical filaments are observed \cite{Zealetal80} in places of the 
jets termination, they radiate in HI, [OI], [NII], [SII] lines.  
Optical filaments in the bipolar nebula are located at $\pm 0.5^{\circ}$
or $\pm 50\,pc$ from SS433. A total
energy of the nebula is $E_{k} \sim 2\cdot10^{51}\, erg$ \cite{Zealetal80},
which corresponds to the jet kinetic luminosity 
$L_k \sim 3\cdot10^{39}\,erg/s$
for 20000 years. The observed velocity dispersion in the filaments is
$\sim 50 \,km/s$, however [NII]/H$\alpha$ line ratio corresponds to
dispersion $\sim 300 \,km/s$ \cite{Zealetal80}. SS433 is an edge-on system,
$i = 79^{\circ}$. If one takes into account this factor,
the velocity dispersion may reach $250-300 \,km/s$. 
The ULXs nebulae studied by us, have about the same sizes, the same
line luminosities and about the same total energy budget  
$10^{51-52}$~erg \cite{FaAbSh06}.

%Fig1
\begin{figure}
\centering
\includegraphics[height=3.5cm]{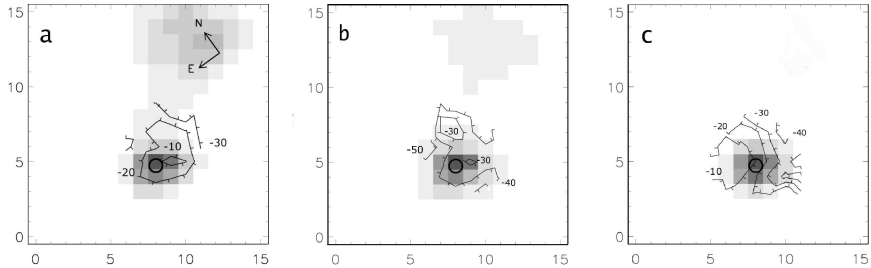}
\caption{H$\alpha$ (a), [SII]\,$\lambda6717, 6734$ (b) and
[OIII]\,$\lambda4959, 5007$ (c)
$15\times15"$ maps of MF\,16 nebula surrounding the NGC\,6946 ULX-1.
Circles show location of the X-ray source from GHANDRA's data.
Equal radial velocty lines are shown. Marks show a direction of 
increasing of the velocity absolute value.
}
\label{fig:1}
\end{figure}

In Fig\,1 we present emission line maps of the nebula MF\,16
("a peculiar SN remnant"), surrounding ULXs in the galaxy 
NGC\,6946. In [OIII]\,$\lambda 4959, 5007$ lines we found a radial velocity 
gradient across the nebula in East-West direction. In later observations
using spectrograph SCORPIO \cite{AbFaSh06} in a long-slit mode we confirmed
this gradient. The radial velocity gradient reaches 100\,km/s across the
whole nebula (20\,pc).

%Fig2
\begin{figure}
\centering
\includegraphics[height=3.8cm]{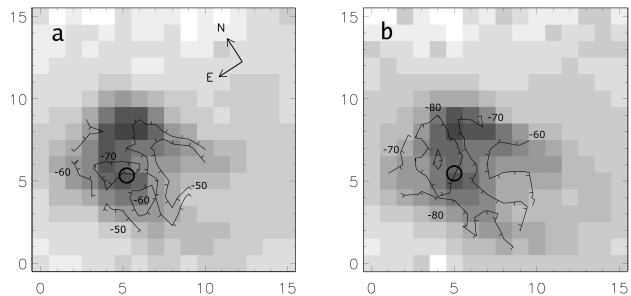}
\caption{H$\alpha$ (a) and [SII]\,$\lambda6717, 6734$ (b) maps
of the nebula surrounding the
IC\,342 ULX-1. Designation are the same as in Fig\,1.
}
\label{fig:2}
\end{figure}

In Fig\,2 we present results of observations of the nebula surrounding
the ULX-1 in the galaxy IC\,342 using the same device. In this nebulae 
we have also detected a radial velocity gradient as a whole expansion of
the nebula $\pm 20$\,km/s. This nebula is not bright because of strong
light absorption in direction of IC\,342.

%Fig3
\begin{figure}
\centering
\includegraphics[height=4cm]{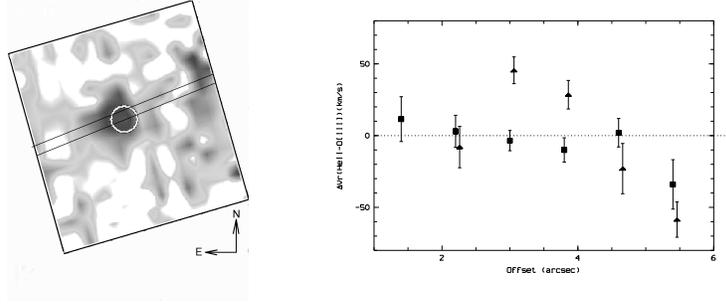}
\caption{HeII\,$\lambda\,4686$ map of the nebula surrounding
the Holmberg\,II ULX-1 (left). Slit positions in LS-spectroscopy are
also shown. Radial velocities (right) of this line measured along the slits
relative to the [OIII]\,$\lambda\,5007$ line are shown 
by squares for the upper slit and by triangles for the bottom slit.
}
\label{fig:3}
\end{figure}

Fig\,3 presents results of observations of Holmberg II ULX-1 nebula 
with the MPFS and SCORPIO spectrograps. Bright He\,II\,$\lambda 4686$ 
emission line is observed in this nebula. We have also detected a radial 
velocity gradient $\pm 50$\,km/s on a spatial scale $\pm 50$\,pc in
the nebula \cite{Leh_ea05}. The radial velocity across the nebula has been
measured in more details in observations with SCORPIO.    

In all three studied ULXs nebulae diagnostic line ratios indicate
collisional excitation of the gas (see \cite{AbFaSh06} for more details). 
We obtain two more conclusions which may be principal for understanding of the
ULXs: \\
1. The radial velocity gradient of $50-100$\,km/s on spatial scales
$20-100$\,pc testify that the nebulae are dynamically perturbed. The IMBHs
can not perturb the interstellar gas on such big scales, the capture Bondi 
radii are not greater than 0.1\,pc. These nebulae can not be SNRs,
they are too big and energetic, they do not satisfy to standard relations for 
SNRs. It is very probable that the nebulae are powered by stellar wind
or jets like that it is in SS\,433.\\
2. For explanation of luminosities in high excitation lines and of the
whole spectrum we need an additional source of hard UV radiation
\cite{AbFaSh06}. The source luminosity is the same huge 
($\sim 10^{40}$~erg/s) as that in X-rays.

%Fig4
\begin{figure}
\centering
\includegraphics[width=11.7cm]{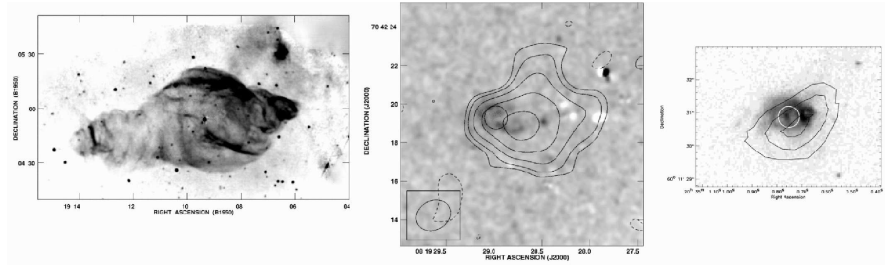}
\caption{
Three nebulae in the same scale in parsecs. VLA image of W\,50
\cite{Dubetal98} with SS\,433 in center, left; Holmberg\,II ULX-1
in HST HeII image with VLA isophotes, middle \cite{MiMuNe05} and
NGC\,6946 ULX-1 in HST H$\alpha$+[SII] image with VLA isophotes
\cite{vanDyketal94}, right. Circles show X-ray Chandra positions.
}
\label{fig:4}
\end{figure}

In Fig.\,4 we present in the same linear scale the nebula W\,50 together 
with nebulae surrounging ULXs in Holmberg\,II and NGC6946 galaxies 
\cite{Leh_ea05}, \cite{FaAbSh06}. The nebulae in Holmberg\,II and NGC\,6946 
have circle-like features in the line-images. In both cases the radio sources 
are shifted to a brighter circle-like feature. The radio sources are 
not resolved. In the both cases the part of the nebulae coinciding with 
radio source is approaching, the opposite part is receding \cite{Leh_ea05},
\cite{FaAbSh06}. At some imagination one may conclude that the nebulae
around these two ULXs are face-on versions ($i=10^{\circ} - 30^{\circ}$)
of the SS\,433 nebula. We need to continue observations to take more
representative sample of nebulae connected with ULXs.

\section*{Acknowledgements}
This work has been supported by the RFBR grant
04--02--16349 and  RFBR/JSPC grant N\, 05--02--19710. 
The authors are grateful to the SOC and LOC of the
Workshop "Science Perspectives for 3D Spectroscopy" for support.

%
% BibTeX users please use
\bibliographystyle{}
\bibliography{}
%
% Non-BibTeX users please follow the syntax
% the syntax of "referenc.tex" for your own citations

\printindex
\end{document}